\documentclass[reprint,superscriptaddress,amsmath,amssymb,aps]{revtex4-1}
\usepackage[utf8]{inputenc}
\usepackage{physics}
\usepackage{graphicx}
\usepackage{siunitx}
\usepackage{soul,color}
\usepackage{float}

\graphicspath{ {Images/} }

\begin{document}
\title{Calculating Opacity in Hot, Dense Matter using Second-Order Electron-Photon and Two-Photon Transitions to Approximate Line Broadening}

\newcommand{\Phys}
{Plasma Physics Group, Blackett Laboratory, Imperial College London, London, SW7 2AZ, UK}

\author{R.A.~Baggott}
\email{r.baggott@imperial.ac.uk}
\affiliation{\Phys}

\author{S.J.~Rose} 
\affiliation{\Phys}

\author{S.P.D.~Mangles} 
\affiliation{\Phys}

\begin{abstract}
Calculations of the opacity of hot, dense matter require models for plasma line broadening.  However, the most general theories are too complex to calculate directly and some approximation is inevitably required.  The most widely-used approaches focus on the line centre, where a Lorentzian shape is obtained.  Here, we demonstrate that in the opposite limit, far from the line centre, the opacity can be expressed in terms of second-order transitions, such as electron-photon and two-photon processes.  We suggest that this insight could form the basis for a new approach to improve calculations of opacity in hot, dense matter.  Preliminary calculations suggest that this approach could yield increased opacity away from absorption lines.
\end{abstract}

\maketitle

The opacity of hot, high density material is important for both terrestrial and astrophysical applications, for example in modelling inertial confinement fusion \cite{dittrich_design_2014, hu_first-principles_2014, hill_modelling_2012}, stellar physics \cite{woosley_evolution_2002, blancard_solar_2012} and planetary interiors \cite{saumon_shock_2004, guillot_interiors_2005, swift_mass-radius_2012}.  Predicted opacities, alongside equations of state and the abundances of heavy elements, are key inputs to solar models \cite{asplund_chemical_2009}.  An increase in predicted opacities could help to reconcile discrepancies between solar models and helioseismic measurements that presently exist \cite{basu_helioseismology_2008}.  

Some recent measurements of iron under solar conditions have yielded opacities in excess of predictions \cite{bailey_higher-than-predicted_2015,nagayama_systematic_2019}.  In particular, increased opacity was observed in regions between characteristic absorption lines.  Such regions make the strongest contribution to the Rosseland mean opacity, which encapsulates the influence of the opacity on radiation transport \cite{huebner_opacity_2014}.  However, the measured opacities cannot be explained within the current theoretical consensus \cite{badnell_updated_2005,iglesias_updated_1996} and are hard to reconcile with the upper limit for single-photon dipole absorption imposed by f-sum rules \cite{iglesias_enigmatic_2015}.

Such experimental findings have prompted recent interest in two-photon transitions as a possible source of opacity.  Two-photon transitions are obtained from second-order perturbation theory and can be interpreted as the simultaneous absorption or emission of two photons.  Since the energy required for the transition may be divided between the two photons in any ratio, two-photon processes yield a continuum of absorption reminiscent of the additional opacity seen experimentally.  Furthermore, being at second-order in perturbation theory, two-photon opacity may exceed the limits of f-sum rules.  However, published results have reached mixed conclusions, with the most detailed calculations finding that two-photon rates should be negligible under the relevant conditions \cite{more_opacity_2017, pain_note_2018, kruse_two-photon_2019}.

Two-photon processes are not the only mechanism unconstrained by the f-sum rule. In dense plasmas, the absorbing ions are also perturbed by the plasma electrons and ions.  This results in broadening of the spectral lines, a process that has been widely explored \cite{griem_stark_1959, baranger_general_1958, dufty_charge-density_1969, lee_plasma_1988}.  We can, however, describe line broadening in a novel way by identifying a process that is the collisional analogue of a two-photon transition.  This process, which we call the electron-photon process, consists of a photon absorption and an electron collision occurring simultaneously.  Although line broadening is not usually approached in this way, the appearance of some satellite lines has been explained in terms of electron-photon transitions \cite{baranger_light_1961,chappell_non-thermal_1970}.

Both the electron-photon description and more conventional line broadening approaches aim to describe the same physical process.  However, the extent to which they are formally equivalent is not immediately apparent.  A similar uncertainty exists in the case of two-photon transitions.  While it has been suggested that two-photon processes might be interpreted as line broadening by background radiation fields \cite{more_semiclassical_1994}, this relationship remains relatively unexplored.

In this work, we demonstrate that we can approximate line broadening in the far wings of the lines with the electron-photon cross-section. The formalism developed for electron broadening is easily extended to the case of broadening by radiation. This allows us to give an expression for broadening due to photons, which can be approximated by the two-photon cross-section in the line wings.  This treatment suggests a new perspective on opacity that may allow improved calculations in spectral regions away from absorption lines, which are highly relevant to stellar modelling.

We begin by considering line broadening by collisions and its relationship to the second-order electron-photon process.  The overall lineshape is due to both the electrons and ions.  It is often assumed that the ions can be treated statically, while the electrons make both a static and dynamic contribution. The lineshape can then be written as \cite{griem_spectral_1974, iglesias_excited_2010}:
\begin{equation}
    I \left( \omega \right) = \int \dd{\vb{F}} W (\vb{F}) J (\omega, \vb{F} ) ,
\end{equation}
with
\begin{multline}
    \label{eq:line_shape}
    J(\omega, \vb{F}) = -\frac{1}{\pi} \Im \sum_{a,b,c,d} \mel{b}{\vb{\epsilon}\vdot\vb{r}}{a}\mel{c}{\vb{\epsilon}\vdot\vb{r}}{d} \\
    \times \mel{ab}{\left[ \Delta\omega - \vb{r}\vdot\vb{F} \middle/ \hbar  - \phi\left(\omega\right)\right]^{-1}\rho}{cd} .
\end{multline}
Here, $\vb{F}$ represents the static microfield, which introduces a line shift. This becomes a line broadening due to integration over the statistical distribution, $W(\vb{F})$.  Since this work is concerned with broadening by electron collisions, we can neglect $\vb{F}$ for brevity without loss of generality.  The sum runs over the atomic states, both bound and continuum.  The density operator, $\rho$, gives the statistical populations of these states.  Notably, this expression does not necessarily reduce to a form that would be bound by the f-sum rule.

In order to make a comprehensive comparison with the second-order electron-photon process, it is necessary to start with a general form for the broadening operator, $\phi(\omega)$, where the matrix elements are given by \cite{iglesias_electron_2005}
\begin{multline}
    \label{eq:line_mat_elems}
    \phi_{ab,cd}(\omega) = \frac{1}{\hbar^2} \int\frac{\dd{\vb{k}}}{(2\pi)^3} \int_{-\infty}^{\infty}\frac{\dd{\Omega}}{2\pi} S(k,\Omega) \\ \times \left( \sum_e \delta_{bd} \frac{V_{ae}(\vb{k})V_{ec}(\vb{-k})}{\Delta\omega_{eb} - \Omega + i\eta} \rho_e \rho_c^{-1} + \sum_e \delta_{ac} \frac{V_{de}(\vb{-k})V_{eb}(\vb{k})}{\Delta\omega_{ae} - \Omega + i\eta} \right. \\
    \left. - V_{ac}(\vb{k})V_{db}(\vb{-k}) \left[ \frac{1}{\Delta\omega_{cb} - \Omega + i\eta} + \frac{\rho_a \rho_c^{-1}}{\Delta\omega_{ad} - \Omega + i\eta}\right] \right).
\end{multline}
We define $\omega$ such that positive values correspond to absorption.  The frequency detuning is then defined as $\Delta\omega_{ab} = (\omega_b - \omega_a) - \omega$.  $V_{ij}(\vb{k}) = \mel{i}{V(\vb{k})}{j}$ are the matrix elements of the electron-radiator interaction \footnote{The perturbing electron interacts simultaneously with the bound electron and nucleus, so that the overall interaction consists of dipole and higher order terms}.  The dynamic structure factor of the plasma electrons, $S(k,\Omega)$, encompasses much of the physics of the collisions, which can alternatively be thought of in terms of absorption or emission of plasmons.

We first consider the case where we are far from the line center.  Making an expansion of Eq.~\ref{eq:line_shape} in orders of ${\phi}/{\Delta\omega}$, we obtain \cite{chappell_non-thermal_1970, iglesias_excited_2010}
\begin{multline}
    \label{eq:line_exp}
    J\left(\omega\right) = - \frac{1}{\pi} \Im \sum_{a,b,c,d} \mel{b}{\left(\vb{\epsilon}\vdot\vb{r}\right)}{a} \mel{c}{ \left(\vb{\epsilon}\vdot\vb{r}\right)}{d} \\ \times
    \mel{ab}{\left(\frac{1}{\Delta\omega} + \frac{1}{\Delta\omega}\phi(\omega)\frac{1}{\Delta\omega} - \mathcal{O}(\Delta\omega^{-3}) \right)\rho}{cd} .
\end{multline}
Introducing Eq.~\ref{eq:line_mat_elems}, the lowest order contribution in this expansion is then given by
\begin{align}
    \label{eq:line_shape_full}
    J(\omega) \approx &-\frac{1}{\pi} \sum_{a,b,c,d} \mel{b}{\left(\vb{\epsilon}\vdot\vb{r}\right)}{a} \mel{c}{ \left(\vb{\epsilon}\vdot\vb{r}\right)}{d} \frac{\Im \phi_{ab,cd}(\omega)}{\Delta\omega_{ab}\Delta\omega_{cd}} \nonumber \\
    \approx &\frac{1}{\hbar^2} \int \frac{\dd{\vb{k}}}{(2\pi)^3} \sum_{a,b,c,d,e} \bigg[ \nonumber \\
     &\frac{d_{ba} d_{cb} V_{ae}(\vb{k})V_{ec}(\vb{-k})}{\Delta\omega_{ab}\Delta\omega_{cb}}
    S(k,\Delta\omega_{eb}) \rho_e  \nonumber \\
    +&\frac{d_{bc} d_{cd}V_{de}(\vb{-k})V_{eb}(\vb{k})}{\Delta\omega_{cb}\Delta\omega_{cd}}
    S(k,\Delta\omega_{ce}) \rho_c \nonumber \\
    -&\frac{d_{ba} d_{cd}V_{ac}(\vb{k})V_{db}(\vb{-k})}{\Delta\omega_{ab}\Delta\omega_{cd}}
    S(k,\Delta\omega_{cb}) \rho_c \nonumber \\
    - &\frac{d_{ba} d_{cd}V_{ac}(\vb{-k})V_{db}(\vb{k})}{\Delta\omega_{ab}\Delta\omega_{cd}}
    S(k,\Delta\omega_{ad}) \rho_a \bigg] ,
\end{align}
where we have abbreviated the dipole matrix elements according to $d_{ij} = \mel{i}{\vb{\varepsilon}\vdot\vb{r}}{j}$.  We now compare this with the rate for an electron-photon transition, calculated using second-order perturbation theory \cite{rahman_high-energy_1978, murillo_dense_1998}:
\begin{multline}
    w_{e \gamma} = \frac{2\pi}{\hbar} \int \frac{\dd{\vb{k}}}{(2\pi)^3} S(k,\omega_{e})
    \pi e^2 E_0^2(\omega_{\gamma}) \\ \times
    \abs{ \sum_n \frac{\mel{f}{\vb{\epsilon}\vdot\vb{r}}{n}\mel{n}{V(\vb{k})}{i}}{E_n - E_i - \hbar\omega_{e}} + \left(e \leftrightarrow \gamma\right)}^2 \\ \times 
    \delta \left( E_f - E_i -\hbar\omega_{e} - \hbar\omega_{\gamma} \right),
\end{multline}
where the positions of the electron and photon are exchanged in the second term.  We can use the delta function to express the collision energy in terms of the frequency detuning:
\begin{equation}
    \omega_{e} = (\omega_f - \omega_i) - \omega_{\gamma} = \Delta\omega_{if} .
\end{equation}
Now taking the total opacity due to electron-photon transitions, and expanding the square modulus, we can obtain a quantity corresponding to $J(\omega)$ of Eq.~\ref{eq:line_shape_full}:
\begin{align}
    \label{eq:electron-photon_full}
    J(\omega) \approx &\frac{\hbar}{2 \pi^2 e^2 E_0^2(\omega_\gamma)}\sum_{i,f} w_{e \gamma} \rho_i \nonumber \\
    \approx &\frac{1}{\hbar^2} \int \frac{\dd{\vb{k}}}{(2\pi)^3} \sum_{i,f,m,n} \bigg[
    \frac{d_{mf}d_{fn} V_{ni}(\vb{k})V_{im}(\vb{-k})}{\Delta\omega_{mf}\Delta\omega_{nf}} \nonumber \\
    +&\frac{d_{im}d_{ni}V_{mf}(\vb{-k})V_{fn}(\vb{k})}{\Delta\omega_{im}\Delta\omega_{in}}
    -\frac{d_{im}d_{fn}V_{ni}(\vb{k})V_{mf}(\vb{-k})}{\Delta\omega_{im}\Delta\omega_{nf}} \nonumber \\
    - &\frac{d_{mf}d_{ni}V_{im}(\vb{-k})V_{fn}(\vb{k})}{\Delta\omega_{mf}\Delta\omega_{in}}
     \bigg]S(k,\Delta\omega_{if}) \rho_i.
\end{align}
Comparing Eq.~\ref{eq:line_shape_full} and Eq.~\ref{eq:electron-photon_full}, we see that, with some relabelling of summation variables, the two expressions are identical.  There is therefore exact equivalence between the electron-photon process and the first term in the line shape expansion. Thus, in the far line wings the broadened transition can be expressed in terms of electron-photon transitions.  Closer to the line center, it will become necessary to include further terms in the expansion.  It is possible that such terms correspond to higher-order transitions involving a photon together with two or more electron collisions.

We can also consider line broadening by background radiation.  The broadening operator can be expressed in terms of the Fourier transform of the electric field autocorrelation \cite{dufty_charge-density_1969}.  In the radiation case, we then apply the Wiener-Khinchin theorem, which relates the autocorrelation function of the fluctuating radiation field to the spectral energy density:
\begin{equation}
    \int_{-\infty}^{\infty} \dd{t} e^{ i\omega t} \langle E(\tau) E(\tau+ t)\rangle = u(\omega) .
\end{equation}
Using this relation, we can then derive a broadening operator, analogous to Eq.~\ref{eq:line_mat_elems}, for broadening by background radiation: 
\begin{multline}
\label{eq:line_mat_elems_rad}
    \phi_{ab,cd} = \frac{1}{\hbar^2} \int_{-\infty}^{\infty}\frac{\dd{\Omega}}{2\pi} 2 \pi e^2 u(\Omega) \\ \times \left( \sum_e \delta_{bd} \frac{d_{ae}d_{ec}}{\Delta\omega_{eb} - \Omega - i\eta} \rho_e \rho_c^{-1} + \sum_e \delta_{ac} \frac{d_{de}d_{eb}}{\Delta\omega_{ae} - \Omega - i\eta} \right. \\
    \left. - d_{ac}d_{db} \left[ \frac{1}{\Delta\omega_{cb} - \Omega - i\eta} + \frac{\rho_a \rho_c^{-1}}{\Delta\omega_{ad} - \Omega - i\eta}\right] \right) .
\end{multline}
Substituting this into Eq.~\ref{eq:line_exp}, just as in the collisional case, we can obtain the two-photon rate as the lowest-order contribution,
\begin{multline}
\label{eq:two_photon_rate}
    J(\omega) \approx \frac{1}{\hbar^2} \sum_{i,f,m,n} \bigg[
    \frac{d_{mf}d_{fn} d_{ni}d_{im}}{\Delta\omega_{mf}\Delta\omega_{nf}} 
    +\frac{d_{im}d_{ni}d_{mf}d_{fn}}{\Delta\omega_{im}\Delta\omega_{in}} \\
    -\frac{d_{im}d_{fn}d_{ni}d_{mf}}{\Delta\omega_{im}\Delta\omega_{nf}} 
    -\frac{d_{mf}d_{ni}d_{im}d_{fn}}{\Delta\omega_{mf}\Delta\omega_{in}}
     \bigg] 2 \pi e^2 u(\Delta\omega_{if}) \rho_i.
\end{multline}
Subsequent terms in the expansion possibly correspond to higher-order multiphoton processes.  With an appropriate definition of the energy density, $u(\omega)$, the expansion includes not only stimulated two-photon processes but also processes incorporating spontaneous emission.

In principle then, a sufficiently general treatment of the line broadening would already include opacity due to second-order processes.  This is the case even for second-order transitions with no directly intermediate state, such as the $1s-2s$ case.  The second-order opacity here arises from the far line wings of the $1s-2p$ (and higher $np$) line, which extend into the energy range between $1s$ and $2s$.  Such transitions had previously been thought of as a `pure' two-photon contribution, with no equivalent in one-photon approaches \cite{goldstein_radiative_1991}.

In practice however, the general broadening operators given in Eqs. \ref{eq:line_mat_elems} and \ref{eq:line_mat_elems_rad} are computationally intractable.  As a result, conventional approaches to opacity calculations depend on approximations that are valid close to the line center \footnote{A more general treatment may be retained in detailed calculations of individual line shapes, however these are not tractable for the large number of lines present in opacity calculations.}.  In the isolated line limit, which requires $\Delta\omega \to 0$ for only the line in question \cite{cooper_broadening_1967}, the line shape has the familiar Lorentzian form
\begin{equation}
    \label{eq:lorentzian}
    J(\omega) = -\frac{1}{\pi} \sum_{a,b} \abs{d_{ab}}^2 \frac{\gamma_{ab}}{\Delta\omega_{ab}^2 + \gamma_{ab}^2} \rho_{a} .
\end{equation}
In this form, the opacity can be constrained using the f-sum rule.  The width is given by the diagonal components of the broadening operator \cite{baranger_general_1958, iglesias_electron_2005}
\begin{align}
    \label{eq:impact}
    \gamma_{ab} &= \Im \phi_{ab,ab}  \nonumber \\
    &=  \left\langle\frac{n_e v}{2}  \left[  \sigma_a + \sigma_b + \int \dd{\Omega}\left|F_a(\Omega) - F_b(\Omega) \right|^2 \right] \right\rangle_\text{Av} .
\end{align}
Here, $F$ are the contributions due to elastic collisions.  The $\Delta\omega \to 0$ limit may be relaxed somewhat for this contribution, for example using the approach of Lee \cite{lee_plasma_1988}. The $\sigma$ are the inelastic contribution, due to collisional and radiative rates.  In particular, this includes the rate of spontaneous radiative emission, which leads to what is often referred to as the `natural' width of the state.  This interpretation only holds in the line centre.  In the line wings, the concept of natural width should give way to two-photon transitions involving spontaneous emission.

To summarize, the most general theory for line broadening is not constrained by the f-sum rule.  However, the approximation commonly used in opacity calculations (Eq.~\ref{eq:lorentzian}) preserves the f-sum rule.  We can improve upon this by including off-diagonal terms of the broadening operator that are not included in Eq.~\ref{eq:lorentzian}. These off-diagonal terms can arise either from collisions with plasma electrons or from a second photon. In our treatment, we approximate these off-diagonal terms with the electron-photon and two-photon cross-sections (Eqs.~\ref{eq:electron-photon_full}~and~\ref{eq:two_photon_rate}).

Employing this approximation in the line wings opens up the possibility for a new approach to opacity calculations based on two-photon and electron-photon processes.  This approach would be most accurate in the regions between lines.  Since these regions contribute strongly to the Rosseland mean opacity, this new approach would be particularly suited to opacity calculations for radiative transport.

\begin{figure}
    \centering
    \includegraphics[width=\linewidth]{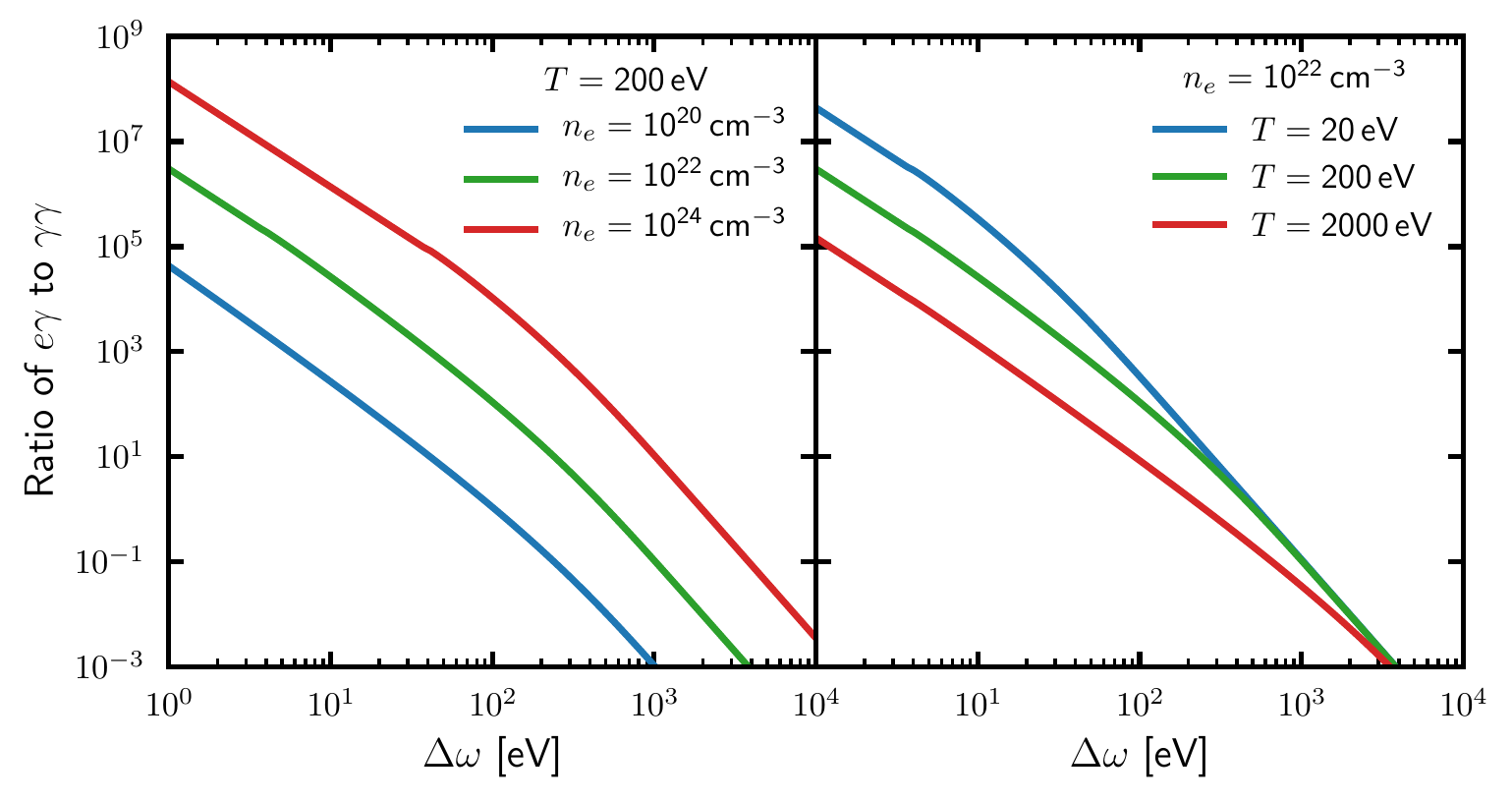}
    \caption{The ratio of electron-photon to two-photon processes shown against detuning from the line center for typical solar conditions.  The rates have been calculated in thermal equilibrium, using the RPA structure factor, which will be valid for a weakly-coupled plasma \cite{kremp_quantum_2005}.}
    \label{fig:ratio_plot}
\end{figure}

\begin{figure}
    \centering
    \includegraphics[width=\linewidth]{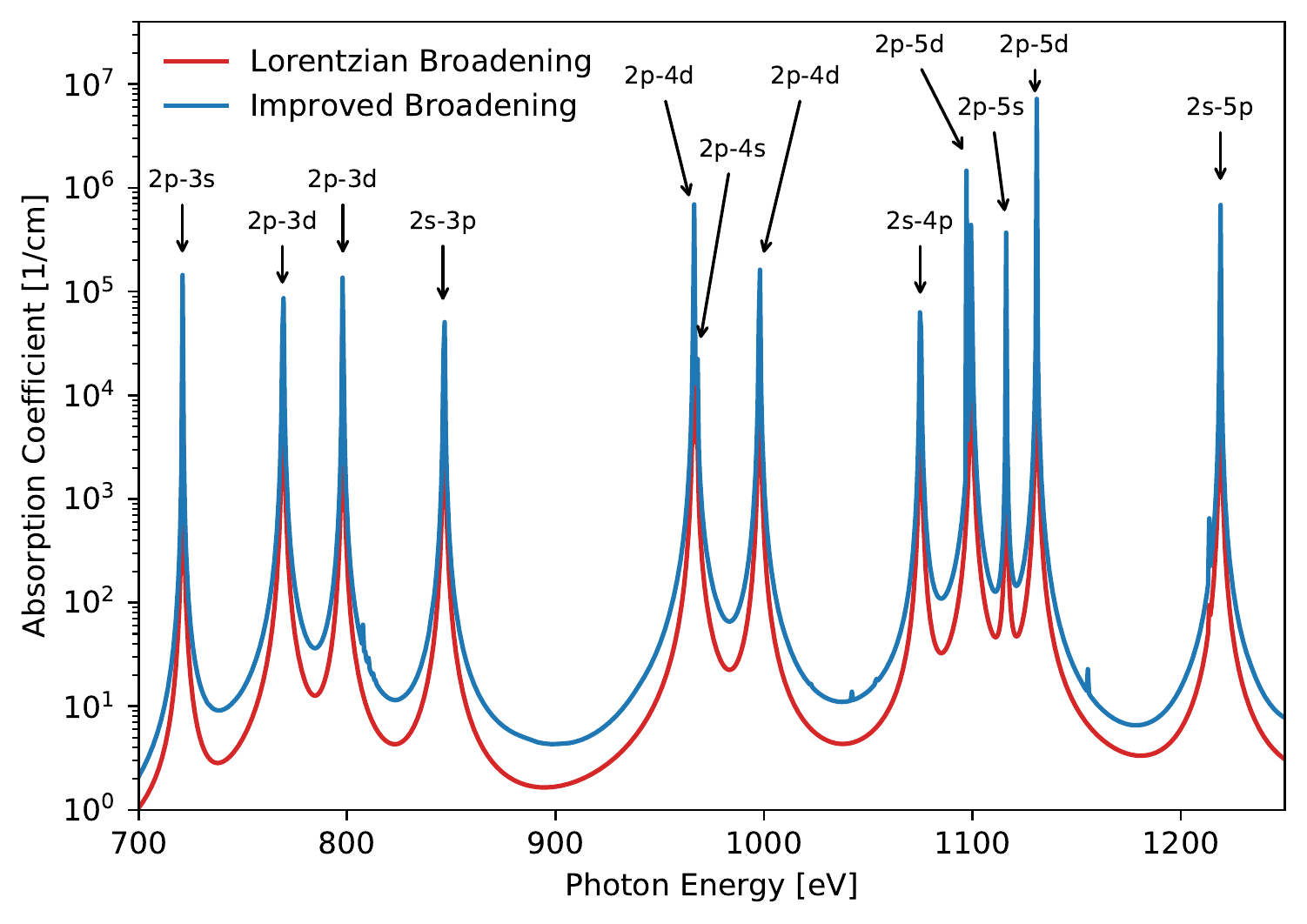}
    \caption{Calculated opacity for Ne-like iron in a plasma with $T=\SI{180}{eV}$ and $n_e = \SI{1.88e21}{cm^{-3}}$.  The simplified Lorentzian broadening (Eq.~\ref{eq:lorentzian}) is compared with an improved broadening using the electron-photon approximation (Eq.~\ref{eq:electron-photon_full}). The same atomic data have been used in both cases.}
    \label{fig:ne-like}
\end{figure}

As a first step towards carrying out opacity calculations in a second-order picture, we can determine the relative importance of electron-photon and two-photon processes.  If both collisions and radiation are treated in the dipole approximation, then, by comparing Eqs.~\ref{eq:electron-photon_full} and \ref{eq:two_photon_rate}, the ratio of the two processes can be written simply as a function of the detuning:
\begin{equation}
	\frac{w_{e\gamma}}{w_{\gamma\gamma}} = \int \frac{\dd{\vb{k}}}{(2\pi)^3} k^2 \left(\frac{4 \pi e^2}{k^2}\right)^2 S(k, \omega) (2\pi e^2 u(\omega))^{-1} .
\end{equation}
This expression has been evaluated for typical solar conditions in Fig.~\ref{fig:ratio_plot}.  We find that the electron-photon process dominates over the two-photon process for moderate detuning ($\lesssim \SI{100}{eV}$).  This is consistent with calculations suggesting negligible two-photon contributions in the \SIrange[range-phrase=--, range-units=single]{6}{10}{\angstrom} range \cite{kruse_two-photon_2019}.  As a result, we focus initially on the electron-photon process.  However, the two-photon process should become the more important for larger detuning, for example between the K- and L-shells of heavier elements.

A preliminary calculation of the opacity of neon-like iron under solar conditions is shown in Fig.~\ref{fig:ne-like} (see Supplemental Material for technical details \footnote{See Supplemental Material at [url], which includes Refs.~\cite{herbst_quantum_2019, dalgarno_exact_1955, robinson_core-excitation_1998, kremp_quantum_2005}, for technical details of the electron-photon opacity calculation.}).  The opacity calculated using the electron-photon approximation is compared with an equivalent calculation using the conventional, Lorentzian broadening.  For the range shown, the opacity is dominated by L-shell bound-bound opacity.  Where it is valid (i.e., in the regions between the lines), the electron-photon approach leads to an increase in opacity of up to 200\%.  Since radiation transport in hot, dense matter is dominated by these spectral regions, this suggests that the electron-photon approach could have significant impact on opacity calculations for radiation transport applications.  

This result is also promising in respect of recent disagreements between theory and experiment \cite{bailey_higher-than-predicted_2015}.  However, the present preliminary calculation includes only a single charge state.  Reaching a firm conclusion in this regard would require an electron-photon opacity calculation on a par with state-of-the-art conventional codes, incorporating a full distribution of charge states and excited states.

This project has received funding from the European Research Council (ERC) under the European Union's Horizon 2020 research and innovation programme (grant agreement no. 682399).


\begin{thebibliography}{40}%
\makeatletter
\providecommand \@ifxundefined [1]{%
 \@ifx{#1\undefined}
}%
\providecommand \@ifnum [1]{%
 \ifnum #1\expandafter \@firstoftwo
 \else \expandafter \@secondoftwo
 \fi
}%
\providecommand \@ifx [1]{%
 \ifx #1\expandafter \@firstoftwo
 \else \expandafter \@secondoftwo
 \fi
}%
\providecommand \natexlab [1]{#1}%
\providecommand \enquote  [1]{``#1''}%
\providecommand \bibnamefont  [1]{#1}%
\providecommand \bibfnamefont [1]{#1}%
\providecommand \citenamefont [1]{#1}%
\providecommand \href@noop [0]{\@secondoftwo}%
\providecommand \href [0]{\begingroup \@sanitize@url \@href}%
\providecommand \@href[1]{\@@startlink{#1}\@@href}%
\providecommand \@@href[1]{\endgroup#1\@@endlink}%
\providecommand \@sanitize@url [0]{\catcode `\\12\catcode `\$12\catcode
  `\&12\catcode `\#12\catcode `\^12\catcode `\_12\catcode `\%12\relax}%
\providecommand \@@startlink[1]{}%
\providecommand \@@endlink[0]{}%
\providecommand \url  [0]{\begingroup\@sanitize@url \@url }%
\providecommand \@url [1]{\endgroup\@href {#1}{\urlprefix }}%
\providecommand \urlprefix  [0]{URL }%
\providecommand \Eprint [0]{\href }%
\providecommand \doibase [0]{http://dx.doi.org/}%
\providecommand \selectlanguage [0]{\@gobble}%
\providecommand \bibinfo  [0]{\@secondoftwo}%
\providecommand \bibfield  [0]{\@secondoftwo}%
\providecommand \translation [1]{[#1]}%
\providecommand \BibitemOpen [0]{}%
\providecommand \bibitemStop [0]{}%
\providecommand \bibitemNoStop [0]{.\EOS\space}%
\providecommand \EOS [0]{\spacefactor3000\relax}%
\providecommand \BibitemShut  [1]{\csname bibitem#1\endcsname}%
\let\auto@bib@innerbib\@empty
\bibitem [{\citenamefont {Dittrich}\ \emph {et~al.}(2014)\citenamefont
  {Dittrich}, \citenamefont {Hurricane}, \citenamefont {Callahan},
  \citenamefont {Dewald}, \citenamefont {D{\"{o}}ppner}, \citenamefont
  {Hinkel}, \citenamefont {{Berzak Hopkins}}, \citenamefont {{Le Pape}},
  \citenamefont {Ma}, \citenamefont {Milovich}, \citenamefont {Moreno},
  \citenamefont {Patel}, \citenamefont {Park}, \citenamefont {Remington},
  \citenamefont {Salmonson},\ and\ \citenamefont
  {Kline}}]{dittrich_design_2014}%
  \BibitemOpen
  \bibfield  {author} {\bibinfo {author} {\bibfnamefont {T.~R.}\ \bibnamefont
  {Dittrich}}, \bibinfo {author} {\bibfnamefont {O.~A.}\ \bibnamefont
  {Hurricane}}, \bibinfo {author} {\bibfnamefont {D.~A.}\ \bibnamefont
  {Callahan}}, \bibinfo {author} {\bibfnamefont {E.~L.}\ \bibnamefont
  {Dewald}}, \bibinfo {author} {\bibfnamefont {T.}~\bibnamefont
  {D{\"{o}}ppner}}, \bibinfo {author} {\bibfnamefont {D.~E.}\ \bibnamefont
  {Hinkel}}, \bibinfo {author} {\bibfnamefont {L.~F.}\ \bibnamefont {{Berzak
  Hopkins}}}, \bibinfo {author} {\bibfnamefont {S.}~\bibnamefont {{Le Pape}}},
  \bibinfo {author} {\bibfnamefont {T.}~\bibnamefont {Ma}}, \bibinfo {author}
  {\bibfnamefont {J.~L.}\ \bibnamefont {Milovich}}, \bibinfo {author}
  {\bibfnamefont {J.~C.}\ \bibnamefont {Moreno}}, \bibinfo {author}
  {\bibfnamefont {P.~K.}\ \bibnamefont {Patel}}, \bibinfo {author}
  {\bibfnamefont {H.-S.}\ \bibnamefont {Park}}, \bibinfo {author}
  {\bibfnamefont {B.~A.}\ \bibnamefont {Remington}}, \bibinfo {author}
  {\bibfnamefont {J.~D.}\ \bibnamefont {Salmonson}}, \ and\ \bibinfo {author}
  {\bibfnamefont {J.~L.}\ \bibnamefont {Kline}},\ }\href {\doibase
  10.1103/PhysRevLett.112.055002} {\bibfield  {journal} {\bibinfo  {journal}
  {Physical Review Letters}\ }\textbf {\bibinfo {volume} {112}},\ \bibinfo
  {pages} {055002} (\bibinfo {year} {2014})}\BibitemShut {NoStop}%
\bibitem [{\citenamefont {Hu}\ \emph {et~al.}(2014)\citenamefont {Hu},
  \citenamefont {Collins}, \citenamefont {Goncharov}, \citenamefont {Boehly},
  \citenamefont {Epstein}, \citenamefont {McCrory},\ and\ \citenamefont
  {Skupsky}}]{hu_first-principles_2014}%
  \BibitemOpen
  \bibfield  {author} {\bibinfo {author} {\bibfnamefont {S.~X.}\ \bibnamefont
  {Hu}}, \bibinfo {author} {\bibfnamefont {L.~A.}\ \bibnamefont {Collins}},
  \bibinfo {author} {\bibfnamefont {V.~N.}\ \bibnamefont {Goncharov}}, \bibinfo
  {author} {\bibfnamefont {T.~R.}\ \bibnamefont {Boehly}}, \bibinfo {author}
  {\bibfnamefont {R.}~\bibnamefont {Epstein}}, \bibinfo {author} {\bibfnamefont
  {R.~L.}\ \bibnamefont {McCrory}}, \ and\ \bibinfo {author} {\bibfnamefont
  {S.}~\bibnamefont {Skupsky}},\ }\href {\doibase 10.1103/PhysRevE.90.033111}
  {\bibfield  {journal} {\bibinfo  {journal} {Physical Review E}\ }\textbf
  {\bibinfo {volume} {90}},\ \bibinfo {pages} {033111} (\bibinfo {year}
  {2014})}\BibitemShut {NoStop}%
\bibitem [{\citenamefont {Hill}\ and\ \citenamefont
  {Rose}(2012)}]{hill_modelling_2012}%
  \BibitemOpen
  \bibfield  {author} {\bibinfo {author} {\bibfnamefont {E.}~\bibnamefont
  {Hill}}\ and\ \bibinfo {author} {\bibfnamefont {S.}~\bibnamefont {Rose}},\
  }\href {\doibase 10.1016/j.hedp.2012.07.001} {\bibfield  {journal} {\bibinfo
  {journal} {High Energy Density Physics}\ }\textbf {\bibinfo {volume} {8}},\
  \bibinfo {pages} {307} (\bibinfo {year} {2012})}\BibitemShut {NoStop}%
\bibitem [{\citenamefont {Woosley}\ \emph {et~al.}(2002)\citenamefont
  {Woosley}, \citenamefont {Heger},\ and\ \citenamefont
  {Weaver}}]{woosley_evolution_2002}%
  \BibitemOpen
  \bibfield  {author} {\bibinfo {author} {\bibfnamefont {S.~E.}\ \bibnamefont
  {Woosley}}, \bibinfo {author} {\bibfnamefont {A.}~\bibnamefont {Heger}}, \
  and\ \bibinfo {author} {\bibfnamefont {T.~A.}\ \bibnamefont {Weaver}},\
  }\href {\doibase 10.1103/RevModPhys.74.1015} {\bibfield  {journal} {\bibinfo
  {journal} {Reviews of Modern Physics}\ }\textbf {\bibinfo {volume} {74}},\
  \bibinfo {pages} {1015} (\bibinfo {year} {2002})}\BibitemShut {NoStop}%
\bibitem [{\citenamefont {Blancard}\ \emph {et~al.}(2012)\citenamefont
  {Blancard}, \citenamefont {Cossé},\ and\ \citenamefont
  {Faussurier}}]{blancard_solar_2012}%
  \BibitemOpen
  \bibfield  {author} {\bibinfo {author} {\bibfnamefont {C.}~\bibnamefont
  {Blancard}}, \bibinfo {author} {\bibfnamefont {P.}~\bibnamefont {Cossé}}, \
  and\ \bibinfo {author} {\bibfnamefont {G.}~\bibnamefont {Faussurier}},\
  }\href {\doibase 10.1088/0004-637X/745/1/10} {\bibfield  {journal} {\bibinfo
  {journal} {The Astrophysical Journal}\ }\textbf {\bibinfo {volume} {745}},\
  \bibinfo {pages} {10} (\bibinfo {year} {2012})}\BibitemShut {NoStop}%
\bibitem [{\citenamefont {Saumon}\ and\ \citenamefont
  {Guillot}(2004)}]{saumon_shock_2004}%
  \BibitemOpen
  \bibfield  {author} {\bibinfo {author} {\bibfnamefont {D.}~\bibnamefont
  {Saumon}}\ and\ \bibinfo {author} {\bibfnamefont {T.}~\bibnamefont
  {Guillot}},\ }\href {\doibase 10.1086/421257} {\bibfield  {journal} {\bibinfo
   {journal} {The Astrophysical Journal}\ }\textbf {\bibinfo {volume} {609}},\
  \bibinfo {pages} {1170} (\bibinfo {year} {2004})}\BibitemShut {NoStop}%
\bibitem [{\citenamefont {Guillot}(2005)}]{guillot_interiors_2005}%
  \BibitemOpen
  \bibfield  {author} {\bibinfo {author} {\bibfnamefont {T.}~\bibnamefont
  {Guillot}},\ }\href {\doibase 10.1146/annurev.earth.32.101802.120325}
  {\bibfield  {journal} {\bibinfo  {journal} {Annual Review of Earth and
  Planetary Sciences}\ }\textbf {\bibinfo {volume} {33}},\ \bibinfo {pages}
  {493} (\bibinfo {year} {2005})}\BibitemShut {NoStop}%
\bibitem [{\citenamefont {Swift}\ \emph {et~al.}(2012)\citenamefont {Swift},
  \citenamefont {Eggert}, \citenamefont {Hicks}, \citenamefont {Hamel},
  \citenamefont {Caspersen}, \citenamefont {Schwegler}, \citenamefont
  {Collins}, \citenamefont {{N. Nettelmann}},\ and\ \citenamefont
  {Ackland}}]{swift_mass-radius_2012}%
  \BibitemOpen
  \bibfield  {author} {\bibinfo {author} {\bibfnamefont {D.~C.}\ \bibnamefont
  {Swift}}, \bibinfo {author} {\bibfnamefont {J.~H.}\ \bibnamefont {Eggert}},
  \bibinfo {author} {\bibfnamefont {D.~G.}\ \bibnamefont {Hicks}}, \bibinfo
  {author} {\bibfnamefont {S.}~\bibnamefont {Hamel}}, \bibinfo {author}
  {\bibfnamefont {K.}~\bibnamefont {Caspersen}}, \bibinfo {author}
  {\bibfnamefont {E.}~\bibnamefont {Schwegler}}, \bibinfo {author}
  {\bibfnamefont {G.~W.}\ \bibnamefont {Collins}}, \bibinfo {author}
  {\bibnamefont {{N. Nettelmann}}}, \ and\ \bibinfo {author} {\bibfnamefont
  {G.~J.}\ \bibnamefont {Ackland}},\ }\href {\doibase
  10.1088/0004-637X/744/1/59} {\bibfield  {journal} {\bibinfo  {journal} {The
  Astrophysical Journal}\ }\textbf {\bibinfo {volume} {744}},\ \bibinfo {pages}
  {59} (\bibinfo {year} {2012})}\BibitemShut {NoStop}%
\bibitem [{\citenamefont {Asplund}\ \emph {et~al.}(2009)\citenamefont
  {Asplund}, \citenamefont {Grevesse}, \citenamefont {Sauval},\ and\
  \citenamefont {Scott}}]{asplund_chemical_2009}%
  \BibitemOpen
  \bibfield  {author} {\bibinfo {author} {\bibfnamefont {M.}~\bibnamefont
  {Asplund}}, \bibinfo {author} {\bibfnamefont {N.}~\bibnamefont {Grevesse}},
  \bibinfo {author} {\bibfnamefont {A.~J.}\ \bibnamefont {Sauval}}, \ and\
  \bibinfo {author} {\bibfnamefont {P.}~\bibnamefont {Scott}},\ }\href
  {\doibase 10.1146/annurev.astro.46.060407.145222} {\bibfield  {journal}
  {\bibinfo  {journal} {Annual Review of Astronomy and Astrophysics}\ }\textbf
  {\bibinfo {volume} {47}},\ \bibinfo {pages} {481} (\bibinfo {year}
  {2009})}\BibitemShut {NoStop}%
\bibitem [{\citenamefont {Basu}\ and\ \citenamefont
  {Antia}(2008)}]{basu_helioseismology_2008}%
  \BibitemOpen
  \bibfield  {author} {\bibinfo {author} {\bibfnamefont {S.}~\bibnamefont
  {Basu}}\ and\ \bibinfo {author} {\bibfnamefont {H.~M.}\ \bibnamefont
  {Antia}},\ }\href {\doibase 10.1016/j.physrep.2007.12.002} {\bibfield
  {journal} {\bibinfo  {journal} {Physics Reports}\ }\textbf {\bibinfo {volume}
  {457}},\ \bibinfo {pages} {217} (\bibinfo {year} {2008})}\BibitemShut
  {NoStop}%
\bibitem [{\citenamefont {Bailey}\ \emph {et~al.}(2015)\citenamefont {Bailey},
  \citenamefont {Nagayama}, \citenamefont {Loisel}, \citenamefont {Rochau},
  \citenamefont {Blancard}, \citenamefont {Colgan}, \citenamefont {Cosse},
  \citenamefont {Faussurier}, \citenamefont {Fontes}, \citenamefont {Gilleron},
  \citenamefont {Golovkin}, \citenamefont {Hansen}, \citenamefont {Iglesias},
  \citenamefont {Kilcrease}, \citenamefont {MacFarlane}, \citenamefont
  {Mancini}, \citenamefont {Nahar}, \citenamefont {Orban}, \citenamefont
  {Pain}, \citenamefont {Pradhan}, \citenamefont {Sherrill},\ and\
  \citenamefont {Wilson}}]{bailey_higher-than-predicted_2015}%
  \BibitemOpen
  \bibfield  {author} {\bibinfo {author} {\bibfnamefont {J.~E.}\ \bibnamefont
  {Bailey}}, \bibinfo {author} {\bibfnamefont {T.}~\bibnamefont {Nagayama}},
  \bibinfo {author} {\bibfnamefont {G.~P.}\ \bibnamefont {Loisel}}, \bibinfo
  {author} {\bibfnamefont {G.~A.}\ \bibnamefont {Rochau}}, \bibinfo {author}
  {\bibfnamefont {C.}~\bibnamefont {Blancard}}, \bibinfo {author}
  {\bibfnamefont {J.}~\bibnamefont {Colgan}}, \bibinfo {author} {\bibfnamefont
  {P.}~\bibnamefont {Cosse}}, \bibinfo {author} {\bibfnamefont
  {G.}~\bibnamefont {Faussurier}}, \bibinfo {author} {\bibfnamefont {C.~J.}\
  \bibnamefont {Fontes}}, \bibinfo {author} {\bibfnamefont {F.}~\bibnamefont
  {Gilleron}}, \bibinfo {author} {\bibfnamefont {I.}~\bibnamefont {Golovkin}},
  \bibinfo {author} {\bibfnamefont {S.~B.}\ \bibnamefont {Hansen}}, \bibinfo
  {author} {\bibfnamefont {C.~A.}\ \bibnamefont {Iglesias}}, \bibinfo {author}
  {\bibfnamefont {D.~P.}\ \bibnamefont {Kilcrease}}, \bibinfo {author}
  {\bibfnamefont {J.~J.}\ \bibnamefont {MacFarlane}}, \bibinfo {author}
  {\bibfnamefont {R.~C.}\ \bibnamefont {Mancini}}, \bibinfo {author}
  {\bibfnamefont {S.~N.}\ \bibnamefont {Nahar}}, \bibinfo {author}
  {\bibfnamefont {C.}~\bibnamefont {Orban}}, \bibinfo {author} {\bibfnamefont
  {J.-C.}\ \bibnamefont {Pain}}, \bibinfo {author} {\bibfnamefont {A.~K.}\
  \bibnamefont {Pradhan}}, \bibinfo {author} {\bibfnamefont {M.}~\bibnamefont
  {Sherrill}}, \ and\ \bibinfo {author} {\bibfnamefont {B.~G.}\ \bibnamefont
  {Wilson}},\ }\href {\doibase 10.1038/nature14048} {\bibfield  {journal}
  {\bibinfo  {journal} {Nature}\ }\textbf {\bibinfo {volume} {517}},\ \bibinfo
  {pages} {56} (\bibinfo {year} {2015})}\BibitemShut {NoStop}%
\bibitem [{\citenamefont {Nagayama}\ \emph {et~al.}(2019)\citenamefont
  {Nagayama}, \citenamefont {Bailey}, \citenamefont {Loisel}, \citenamefont
  {Dunham}, \citenamefont {Rochau}, \citenamefont {Blancard}, \citenamefont
  {Colgan}, \citenamefont {Coss\'e}, \citenamefont {Faussurier}, \citenamefont
  {Fontes}, \citenamefont {Gilleron}, \citenamefont {Hansen}, \citenamefont
  {Iglesias}, \citenamefont {Golovkin}, \citenamefont {Kilcrease},
  \citenamefont {MacFarlane}, \citenamefont {Mancini}, \citenamefont {More},
  \citenamefont {Orban}, \citenamefont {Pain}, \citenamefont {Sherrill},\ and\
  \citenamefont {Wilson}}]{nagayama_systematic_2019}%
  \BibitemOpen
  \bibfield  {author} {\bibinfo {author} {\bibfnamefont {T.}~\bibnamefont
  {Nagayama}}, \bibinfo {author} {\bibfnamefont {J.~E.}\ \bibnamefont
  {Bailey}}, \bibinfo {author} {\bibfnamefont {G.~P.}\ \bibnamefont {Loisel}},
  \bibinfo {author} {\bibfnamefont {G.~S.}\ \bibnamefont {Dunham}}, \bibinfo
  {author} {\bibfnamefont {G.~A.}\ \bibnamefont {Rochau}}, \bibinfo {author}
  {\bibfnamefont {C.}~\bibnamefont {Blancard}}, \bibinfo {author}
  {\bibfnamefont {J.}~\bibnamefont {Colgan}}, \bibinfo {author} {\bibfnamefont
  {P.}~\bibnamefont {Coss\'e}}, \bibinfo {author} {\bibfnamefont
  {G.}~\bibnamefont {Faussurier}}, \bibinfo {author} {\bibfnamefont {C.~J.}\
  \bibnamefont {Fontes}}, \bibinfo {author} {\bibfnamefont {F.}~\bibnamefont
  {Gilleron}}, \bibinfo {author} {\bibfnamefont {S.~B.}\ \bibnamefont
  {Hansen}}, \bibinfo {author} {\bibfnamefont {C.~A.}\ \bibnamefont
  {Iglesias}}, \bibinfo {author} {\bibfnamefont {I.~E.}\ \bibnamefont
  {Golovkin}}, \bibinfo {author} {\bibfnamefont {D.~P.}\ \bibnamefont
  {Kilcrease}}, \bibinfo {author} {\bibfnamefont {J.~J.}\ \bibnamefont
  {MacFarlane}}, \bibinfo {author} {\bibfnamefont {R.~C.}\ \bibnamefont
  {Mancini}}, \bibinfo {author} {\bibfnamefont {R.~M.}\ \bibnamefont {More}},
  \bibinfo {author} {\bibfnamefont {C.}~\bibnamefont {Orban}}, \bibinfo
  {author} {\bibfnamefont {J.-C.}\ \bibnamefont {Pain}}, \bibinfo {author}
  {\bibfnamefont {M.~E.}\ \bibnamefont {Sherrill}}, \ and\ \bibinfo {author}
  {\bibfnamefont {B.~G.}\ \bibnamefont {Wilson}},\ }\href {\doibase
  10.1103/PhysRevLett.122.235001} {\bibfield  {journal} {\bibinfo  {journal}
  {Phys. Rev. Lett.}\ }\textbf {\bibinfo {volume} {122}},\ \bibinfo {pages}
  {235001} (\bibinfo {year} {2019})}\BibitemShut {NoStop}%
\bibitem [{\citenamefont {Huebner}\ and\ \citenamefont
  {Barfield}(2014)}]{huebner_opacity_2014}%
  \BibitemOpen
  \bibfield  {author} {\bibinfo {author} {\bibfnamefont {W.~F.}\ \bibnamefont
  {Huebner}}\ and\ \bibinfo {author} {\bibfnamefont {W.~D.}\ \bibnamefont
  {Barfield}},\ }\href {\doibase 10.1007/978-1-4614-8797-5} {\emph {\bibinfo
  {title} {Opacity}}},\ \bibinfo {series} {Astrophysics and {Space} {Science}
  {Library}}, Vol.\ \bibinfo {volume} {402}\ (\bibinfo  {publisher} {Springer
  New York},\ \bibinfo {address} {New York, NY},\ \bibinfo {year}
  {2014})\BibitemShut {NoStop}%
\bibitem [{\citenamefont {Badnell}\ \emph {et~al.}(2005)\citenamefont
  {Badnell}, \citenamefont {Bautista}, \citenamefont {Butler}, \citenamefont
  {Delahaye}, \citenamefont {Mendoza}, \citenamefont {Palmeri}, \citenamefont
  {Zeippen},\ and\ \citenamefont {Seaton}}]{badnell_updated_2005}%
  \BibitemOpen
  \bibfield  {author} {\bibinfo {author} {\bibfnamefont {N.~R.}\ \bibnamefont
  {Badnell}}, \bibinfo {author} {\bibfnamefont {M.~A.}\ \bibnamefont
  {Bautista}}, \bibinfo {author} {\bibfnamefont {K.}~\bibnamefont {Butler}},
  \bibinfo {author} {\bibfnamefont {F.}~\bibnamefont {Delahaye}}, \bibinfo
  {author} {\bibfnamefont {C.}~\bibnamefont {Mendoza}}, \bibinfo {author}
  {\bibfnamefont {P.}~\bibnamefont {Palmeri}}, \bibinfo {author} {\bibfnamefont
  {C.~J.}\ \bibnamefont {Zeippen}}, \ and\ \bibinfo {author} {\bibfnamefont
  {M.~J.}\ \bibnamefont {Seaton}},\ }\href {\doibase
  10.1111/j.1365-2966.2005.08991.x} {\bibfield  {journal} {\bibinfo  {journal}
  {Monthly Notices of the Royal Astronomical Society}\ }\textbf {\bibinfo
  {volume} {360}},\ \bibinfo {pages} {458} (\bibinfo {year}
  {2005})}\BibitemShut {NoStop}%
\bibitem [{\citenamefont {Iglesias}\ and\ \citenamefont
  {Rogers}(1996)}]{iglesias_updated_1996}%
  \BibitemOpen
  \bibfield  {author} {\bibinfo {author} {\bibfnamefont {C.~A.}\ \bibnamefont
  {Iglesias}}\ and\ \bibinfo {author} {\bibfnamefont {F.~J.}\ \bibnamefont
  {Rogers}},\ }\href {\doibase 10.1086/177381} {\bibfield  {journal} {\bibinfo
  {journal} {The Astrophysical Journal}\ }\textbf {\bibinfo {volume} {464}},\
  \bibinfo {pages} {943} (\bibinfo {year} {1996})}\BibitemShut {NoStop}%
\bibitem [{\citenamefont {Iglesias}(2015)}]{iglesias_enigmatic_2015}%
  \BibitemOpen
  \bibfield  {author} {\bibinfo {author} {\bibfnamefont {C.~A.}\ \bibnamefont
  {Iglesias}},\ }\href {\doibase 10.1016/j.hedp.2015.03.009} {\bibfield
  {journal} {\bibinfo  {journal} {High Energy Density Physics}\ }\textbf
  {\bibinfo {volume} {15}},\ \bibinfo {pages} {4} (\bibinfo {year}
  {2015})}\BibitemShut {NoStop}%
\bibitem [{\citenamefont {More}\ \emph {et~al.}(2017)\citenamefont {More},
  \citenamefont {Hansen},\ and\ \citenamefont {Nagayama}}]{more_opacity_2017}%
  \BibitemOpen
  \bibfield  {author} {\bibinfo {author} {\bibfnamefont {R.~M.}\ \bibnamefont
  {More}}, \bibinfo {author} {\bibfnamefont {S.~B.}\ \bibnamefont {Hansen}}, \
  and\ \bibinfo {author} {\bibfnamefont {T.}~\bibnamefont {Nagayama}},\ }\href
  {\doibase 10.1016/j.hedp.2017.07.003} {\bibfield  {journal} {\bibinfo
  {journal} {High Energy Density Physics}\ }\textbf {\bibinfo {volume} {24}},\
  \bibinfo {pages} {44} (\bibinfo {year} {2017})}\BibitemShut {NoStop}%
\bibitem [{\citenamefont {Pain}(2018)}]{pain_note_2018}%
  \BibitemOpen
  \bibfield  {author} {\bibinfo {author} {\bibfnamefont {J.-C.}\ \bibnamefont
  {Pain}},\ }\href {\doibase 10.1016/j.hedp.2017.11.004} {\bibfield  {journal}
  {\bibinfo  {journal} {High Energy Density Physics}\ }\textbf {\bibinfo
  {volume} {26}},\ \bibinfo {pages} {23} (\bibinfo {year} {2018})}\BibitemShut
  {NoStop}%
\bibitem [{\citenamefont {Kruse}\ and\ \citenamefont
  {Iglesias}(2019)}]{kruse_two-photon_2019}%
  \BibitemOpen
  \bibfield  {author} {\bibinfo {author} {\bibfnamefont {M.~K.}\ \bibnamefont
  {Kruse}}\ and\ \bibinfo {author} {\bibfnamefont {C.~A.}\ \bibnamefont
  {Iglesias}},\ }\href {\doibase 10.1016/j.hedp.2019.02.004} {\bibfield
  {journal} {\bibinfo  {journal} {High Energy Density Physics}\ }\textbf
  {\bibinfo {volume} {31}},\ \bibinfo {pages} {38} (\bibinfo {year}
  {2019})}\BibitemShut {NoStop}%
\bibitem [{\citenamefont {Griem}\ \emph {et~al.}(1959)\citenamefont {Griem},
  \citenamefont {Kolb},\ and\ \citenamefont {Shen}}]{griem_stark_1959}%
  \BibitemOpen
  \bibfield  {author} {\bibinfo {author} {\bibfnamefont {H.~R.}\ \bibnamefont
  {Griem}}, \bibinfo {author} {\bibfnamefont {A.~C.}\ \bibnamefont {Kolb}}, \
  and\ \bibinfo {author} {\bibfnamefont {K.~Y.}\ \bibnamefont {Shen}},\ }\href
  {\doibase 10.1103/PhysRev.116.4} {\bibfield  {journal} {\bibinfo  {journal}
  {Physical Review}\ }\textbf {\bibinfo {volume} {116}},\ \bibinfo {pages} {4}
  (\bibinfo {year} {1959})}\BibitemShut {NoStop}%
\bibitem [{\citenamefont {Baranger}(1958)}]{baranger_general_1958}%
  \BibitemOpen
  \bibfield  {author} {\bibinfo {author} {\bibfnamefont {M.}~\bibnamefont
  {Baranger}},\ }\href {\doibase 10.1103/PhysRev.112.855} {\bibfield  {journal}
  {\bibinfo  {journal} {Physical Review}\ }\textbf {\bibinfo {volume} {112}},\
  \bibinfo {pages} {855} (\bibinfo {year} {1958})}\BibitemShut {NoStop}%
\bibitem [{\citenamefont {Dufty}(1969)}]{dufty_charge-density_1969}%
  \BibitemOpen
  \bibfield  {author} {\bibinfo {author} {\bibfnamefont {J.~W.}\ \bibnamefont
  {Dufty}},\ }\href {\doibase 10.1103/PhysRev.187.305} {\bibfield  {journal}
  {\bibinfo  {journal} {Physical Review}\ }\textbf {\bibinfo {volume} {187}},\
  \bibinfo {pages} {305} (\bibinfo {year} {1969})}\BibitemShut {NoStop}%
\bibitem [{\citenamefont {Lee}(1988)}]{lee_plasma_1988}%
  \BibitemOpen
  \bibfield  {author} {\bibinfo {author} {\bibfnamefont {R.}~\bibnamefont
  {Lee}},\ }\href {\doibase 10.1016/0022-4073(88)90136-7} {\bibfield  {journal}
  {\bibinfo  {journal} {Journal of Quantitative Spectroscopy and Radiative
  Transfer}\ }\textbf {\bibinfo {volume} {40}},\ \bibinfo {pages} {561}
  (\bibinfo {year} {1988})}\BibitemShut {NoStop}%
\bibitem [{\citenamefont {Baranger}\ and\ \citenamefont
  {Mozer}(1961)}]{baranger_light_1961}%
  \BibitemOpen
  \bibfield  {author} {\bibinfo {author} {\bibfnamefont {M.}~\bibnamefont
  {Baranger}}\ and\ \bibinfo {author} {\bibfnamefont {B.}~\bibnamefont
  {Mozer}},\ }\href {\doibase 10.1103/PhysRev.123.25} {\bibfield  {journal}
  {\bibinfo  {journal} {Physical Review}\ }\textbf {\bibinfo {volume} {123}},\
  \bibinfo {pages} {25} (\bibinfo {year} {1961})}\BibitemShut {NoStop}%
\bibitem [{\citenamefont {Chappell}\ \emph {et~al.}(1970)\citenamefont
  {Chappell}, \citenamefont {Cooper},\ and\ \citenamefont
  {Smith}}]{chappell_non-thermal_1970}%
  \BibitemOpen
  \bibfield  {author} {\bibinfo {author} {\bibfnamefont {W.}~\bibnamefont
  {Chappell}}, \bibinfo {author} {\bibfnamefont {J.}~\bibnamefont {Cooper}}, \
  and\ \bibinfo {author} {\bibfnamefont {E.}~\bibnamefont {Smith}},\ }\href
  {\doibase 10.1016/0022-4073(70)90004-X} {\bibfield  {journal} {\bibinfo
  {journal} {Journal of Quantitative Spectroscopy and Radiative Transfer}\
  }\textbf {\bibinfo {volume} {10}},\ \bibinfo {pages} {1195} (\bibinfo {year}
  {1970})}\BibitemShut {NoStop}%
\bibitem [{\citenamefont {More}(1994)}]{more_semiclassical_1994}%
  \BibitemOpen
  \bibfield  {author} {\bibinfo {author} {\bibfnamefont {R.~M.}\ \bibnamefont
  {More}},\ }in\ \href {\doibase 10.1007/978-1-4899-1576-4_7} {\emph {\bibinfo
  {booktitle} {Laser {{Interactions}} with {{Atoms}}, {{Solids}} and
  {{Plasmas}}}}},\ Vol.\ \bibinfo {volume} {327},\ \bibinfo {editor} {edited
  by\ \bibinfo {editor} {\bibfnamefont {R.~M.}\ \bibnamefont {More}}}\
  (\bibinfo  {publisher} {{Springer US}},\ \bibinfo {address} {{Boston, MA}},\
  \bibinfo {year} {1994})\ pp.\ \bibinfo {pages} {123--161}\BibitemShut
  {NoStop}%
\bibitem [{\citenamefont {Griem}(1974)}]{griem_spectral_1974}%
  \BibitemOpen
  \bibfield  {author} {\bibinfo {author} {\bibfnamefont {H.~R.}\ \bibnamefont
  {Griem}},\ }\href@noop {} {\emph {\bibinfo {title} {Spectral line broadening
  by plasmas}}},\ \bibinfo {series} {Pure and applied physics}\ No.\ \bibinfo
  {number} {v. 39}\ (\bibinfo  {publisher} {Academic Press},\ \bibinfo
  {address} {New York},\ \bibinfo {year} {1974})\BibitemShut {NoStop}%
\bibitem [{\citenamefont {Iglesias}(2010)}]{iglesias_excited_2010}%
  \BibitemOpen
  \bibfield  {author} {\bibinfo {author} {\bibfnamefont {C.~A.}\ \bibnamefont
  {Iglesias}},\ }\href {\doibase 10.1016/j.hedp.2010.01.007} {\bibfield
  {journal} {\bibinfo  {journal} {High Energy Density Physics}\ }\textbf
  {\bibinfo {volume} {6}},\ \bibinfo {pages} {318} (\bibinfo {year}
  {2010})}\BibitemShut {NoStop}%
\bibitem [{\citenamefont {Iglesias}(2005)}]{iglesias_electron_2005}%
  \BibitemOpen
  \bibfield  {author} {\bibinfo {author} {\bibfnamefont {C.~A.}\ \bibnamefont
  {Iglesias}},\ }\href {\doibase 10.1016/j.hedp.2005.08.003} {\bibfield
  {journal} {\bibinfo  {journal} {High Energy Density Physics}\ }\textbf
  {\bibinfo {volume} {1}},\ \bibinfo {pages} {42} (\bibinfo {year}
  {2005})}\BibitemShut {NoStop}%
\bibitem [{Note1()}]{Note1}%
  \BibitemOpen
  \bibinfo {note} {The perturbing electron interacts simultaneously with the
  bound electron and nucleus, so that the overall interaction consists of
  dipole and higher order terms}\BibitemShut {NoStop}%
\bibitem [{\citenamefont {Rahman}\ and\ \citenamefont
  {Faisal}(1978)}]{rahman_high-energy_1978}%
  \BibitemOpen
  \bibfield  {author} {\bibinfo {author} {\bibfnamefont {N.~K.}\ \bibnamefont
  {Rahman}}\ and\ \bibinfo {author} {\bibfnamefont {F.~H.~M.}\ \bibnamefont
  {Faisal}},\ }\href@noop {} {\bibfield  {journal} {\bibinfo  {journal}
  {Journal of Physics B: Atomic and Molecular Physics}\ }\textbf {\bibinfo
  {volume} {11}},\ \bibinfo {pages} {2003} (\bibinfo {year}
  {1978})}\BibitemShut {NoStop}%
\bibitem [{\citenamefont {Murillo}\ and\ \citenamefont
  {Weisheit}(1998)}]{murillo_dense_1998}%
  \BibitemOpen
  \bibfield  {author} {\bibinfo {author} {\bibfnamefont {M.~S.}\ \bibnamefont
  {Murillo}}\ and\ \bibinfo {author} {\bibfnamefont {J.~C.}\ \bibnamefont
  {Weisheit}},\ }\href {\doibase 10.1016/S0370-1573(98)00017-9} {\bibfield
  {journal} {\bibinfo  {journal} {Physics Reports}\ }\textbf {\bibinfo {volume}
  {302}},\ \bibinfo {pages} {1} (\bibinfo {year} {1998})}\BibitemShut {NoStop}%
\bibitem [{\citenamefont {Rose}\ and\ \citenamefont
  {More}(1991)}]{goldstein_radiative_1991}%
  \BibitemOpen
  \bibfield  {author} {\bibinfo {author} {\bibfnamefont {S.}~\bibnamefont
  {Rose}}\ and\ \bibinfo {author} {\bibfnamefont {R.}~\bibnamefont {More}},\
  }in\ \href {https://books.google.co.uk/books?id=nkBPDwAAQBAJ} {\emph
  {\bibinfo {booktitle} {Radiative {Properties} {Of} {Hot} {Dense} {Matter},
  {Proceedings} of the 4th {International} {Workshop}, {Sarasota},
  {Florida}}}},\ \bibinfo {editor} {edited by\ \bibinfo {editor} {\bibfnamefont
  {W.}~\bibnamefont {Goldstein}}, \bibinfo {editor} {\bibfnamefont
  {C.}~\bibnamefont {Hooper}}, \bibinfo {editor} {\bibfnamefont
  {J.}~\bibnamefont {Gauthier}}, \bibinfo {editor} {\bibfnamefont
  {J.}~\bibnamefont {Seely}}, \ and\ \bibinfo {editor} {\bibfnamefont
  {R.}~\bibnamefont {Lee}}}\ (\bibinfo  {publisher} {World Scientific
  Publishing Company},\ \bibinfo {address} {Singapore},\ \bibinfo {year}
  {1991})\BibitemShut {NoStop}%
\bibitem [{Note2()}]{Note2}%
  \BibitemOpen
  \bibinfo {note} {A more general treatment may be retained in detailed
  calculations of individual line shapes, however these are not tractable for
  the large number of lines present in opacity calculations.}\BibitemShut
  {Stop}%
\bibitem [{\citenamefont {Cooper}(1967)}]{cooper_broadening_1967}%
  \BibitemOpen
  \bibfield  {author} {\bibinfo {author} {\bibfnamefont {J.}~\bibnamefont
  {Cooper}},\ }\href {\doibase 10.1103/RevModPhys.39.167} {\bibfield  {journal}
  {\bibinfo  {journal} {Reviews of Modern Physics}\ }\textbf {\bibinfo {volume}
  {39}},\ \bibinfo {pages} {167} (\bibinfo {year} {1967})}\BibitemShut
  {NoStop}%
\bibitem [{\citenamefont {Kremp}\ \emph {et~al.}(2005)\citenamefont {Kremp},
  \citenamefont {Schlanges}, \citenamefont {Kraeft},\ and\ \citenamefont
  {Bornath}}]{kremp_quantum_2005}%
  \BibitemOpen
  \bibfield  {author} {\bibinfo {author} {\bibfnamefont {D.}~\bibnamefont
  {Kremp}}, \bibinfo {author} {\bibfnamefont {M.}~\bibnamefont {Schlanges}},
  \bibinfo {author} {\bibfnamefont {W.-D.}\ \bibnamefont {Kraeft}}, \ and\
  \bibinfo {author} {\bibfnamefont {T.}~\bibnamefont {Bornath}},\ }\href@noop
  {} {\emph {\bibinfo {title} {Quantum {Statistics} of {Nonideal}
  {Plasmas}}}},\ \bibinfo {series} {Springer series on atomic, optical, and
  plasma physics}\ No.~\bibinfo {number} {25}\ (\bibinfo  {publisher}
  {Springer},\ \bibinfo {address} {Berlin},\ \bibinfo {year}
  {2005})\BibitemShut {NoStop}%
\bibitem [{Note3()}]{Note3}%
  \BibitemOpen
  \bibinfo {note} {See Supplemental Material, which includes Refs.~\cite{herbst_quantum_2019, dalgarno_exact_1955, robinson_core-excitation_1998, kremp_quantum_2005}, for technical details of the electron-photon opacity calculation.}\BibitemShut {Stop}%
\bibitem [{\citenamefont {Herbst}\ \emph {et~al.}(2019)\citenamefont {Herbst},
  \citenamefont {Avery},\ and\ \citenamefont {Dreuw}}]{herbst_quantum_2019}%
  \BibitemOpen
  \bibfield  {author} {\bibinfo {author} {\bibfnamefont {M.~F.}\ \bibnamefont
  {Herbst}}, \bibinfo {author} {\bibfnamefont {J.~E.}\ \bibnamefont {Avery}}, \
  and\ \bibinfo {author} {\bibfnamefont {A.}~\bibnamefont {Dreuw}},\ }\href
  {\doibase 10.1103/PhysRevA.99.012512} {\bibfield  {journal} {\bibinfo
  {journal} {Physical Review A}\ }\textbf {\bibinfo {volume} {99}},\ \bibinfo
  {pages} {012512} (\bibinfo {year} {2019})}\BibitemShut {NoStop}%
\bibitem [{\citenamefont {Dalgarno}\ and\ \citenamefont
  {Lewis}(1955)}]{dalgarno_exact_1955}%
  \BibitemOpen
  \bibfield  {author} {\bibinfo {author} {\bibfnamefont {A.}~\bibnamefont
  {Dalgarno}}\ and\ \bibinfo {author} {\bibfnamefont {J.~T.}\ \bibnamefont
  {Lewis}},\ }\href {\doibase 10.1098/rspa.1955.0246} {\bibfield  {journal}
  {\bibinfo  {journal} {Proceedings of the Royal Society of London. Series A.
  Mathematical and Physical Sciences}\ }\textbf {\bibinfo {volume} {233}},\
  \bibinfo {pages} {70} (\bibinfo {year} {1955})}\BibitemShut {NoStop}%
\bibitem [{\citenamefont {Robinson}(1998)}]{robinson_core-excitation_1998}%
  \BibitemOpen
  \bibfield  {author} {\bibinfo {author} {\bibfnamefont {E.~J.}\ \bibnamefont
  {Robinson}},\ }\href {\doibase 10.1103/PhysRevA.58.755} {\bibfield  {journal}
  {\bibinfo  {journal} {Physical Review A}\ }\textbf {\bibinfo {volume} {58}},\
  \bibinfo {pages} {755} (\bibinfo {year} {1998})}\BibitemShut {NoStop}%
\end{thebibliography}
\end{document}


\title{Supplementary Material to: Calculating Opacity in Hot, Dense Matter using Second-Order Electron-Photon and Two-Photon Transitions to Approximate Line Broadening}

\newcommand{\Phys}
{Plasma Physics Group, Blackett Laboratory, Imperial College London, London, SW7 2AZ, UK}

\author{R.A.~Baggott}
\email{r.baggott@imperial.ac.uk}
\affiliation{\Phys}

\author{S.J.~Rose} 
\affiliation{\Phys}

\author{S.P.D.~Mangles} 
\affiliation{\Phys}

\maketitle

\section{Details of Preliminary Opacity Calculation}
The bound state wavefunctions and energies for the calculation presented in Fig. 3 of the main text have been obtained from a Hartree-Fock calculation.  The orbitals have been expanded in a Coulomb-Sturmian basis set \cite{herbst_quantum_2019} with $n_\mathrm{max} = 11$, $l_\mathrm{max} = 3$ and $m_\mathrm{max} = 3$ and an exponent $k = 10$.

Both electron-atom and photon-atom interactions have been modelled in the dipole approximation.  The second-order matrix elements required for the electron-photon opacity are calculated using the Dalgarno-Lewis method \cite{dalgarno_exact_1955}.  This involves calculating an effective intermediate state, $\ket{\Psi^{DL}}$, which satisfies
\begin{equation}
    (E - \hat{H})\ket{\Psi^{DL}} = r\ket{i}
\end{equation}
where $\ket{i}$ is the initial state and $\hat{H}$ is the atomic Hamiltonian.  The two terms corresponding to the photon or electron interacting first have $E=E_i + \hbar\omega$ and $E=E_f - \hbar\omega$ respectively, where $E_i$ and $E_f$ are the energies of the initial and final state.  This can be converted into a matrix equation by expanding the intermediate state in the same basis as the atomic orbitals.  The second-order matrix element is then given by
\begin{equation}
    M = \mel{f}{r}{\Psi^{DL}}
\end{equation}
where $\ket{f}$ is the final state.

The matrix elements have been calculated assuming a single active electron moving in the frozen Hartree-Fock potential of the remaining electrons.  In order to prevent transitions via filled intermediate shells, the components of the intermediate state corresponding to filled orbitals have been projected out according to
\begin{equation}
    \ket{\Psi^{DL}} \rightarrow \ket{\Psi^{DL}} - \sum_\text{k filled} \frac{\mel{k}{r}{i}}{E - E_k} \ket{k} .
\end{equation}
It has been suggested that such transitions should be retained as an approximation of core excitations \cite{robinson_core-excitation_1998}.  However, we found that the transition energies predicted by this approach were not sufficiently accurate, leading to resonances at unphysical energies.  Since excluding these terms can only lower the electron-photon opacity, this does not effect our conclusions.

The dynamic structure factor of the electrons, $S(k,\omega)$ has been calculated using the random phase approximation \cite{kremp_quantum_2005} throughout.